\documentclass[twocolumn,showpacs,preprintnumbers,amsmath,amssymb,prx,superscriptaddress]{revtex4}
\usepackage{mathrsfs}
\usepackage{graphicx}
\usepackage{dcolumn}
\usepackage{bm}
\usepackage{amsmath}
\usepackage{amsfonts}
\usepackage{color}

\begin{document}

\title{A Five Dimensional Generalization of the Topological Weyl Semimetal}
\author{Biao Lian}
\affiliation{Department of Physics, McCullough Building, Stanford University, Stanford, California 94305-4045, USA}
\author{Shou-Cheng Zhang}
\affiliation{Department of Physics, McCullough Building, Stanford University, Stanford, California 94305-4045, USA}

\begin{abstract}
We generalize the concept of three-dimensional topological Weyl semimetal to a class of five dimensional (5D) gapless solids, where Weyl points are generalized to Weyl surfaces which are two-dimensional closed manifolds in the momentum space. Each Weyl surface is characterized by a U(1) second Chern number $C_2$ defined on a four-dimensional manifold enclosing the Weyl surface, which is equal to its topological linking number with other Weyl surfaces in 5D. In analogy to the Weyl semimetals, the surface states of the 5D metal take the form of topologically protected Weyl fermion arcs, which connect the projections of the bulk Weyl surfaces. The further generalization of topological metal in $2n+1$ dimensions carrying the $n$-th Chern number $C_n$ is also discussed.
\end{abstract}

\date{\today}


\maketitle

The study of topological states of matter has generated enormous interest in condensed matter physics in the past decade. Prominent examples of topological states include quantum Hall states, topological insulators and topological superconductors, which all have a bulk gap and possess protected gapless surface states \cite{Hasan2010,Qi2011,Wang2015R,Liu2016}. Recently, the three-dimensional (3D) Weyl semimetal has been studied extensively, which is a gapless topological state that realizes Weyl fermions and the chiral anomaly in a condensed matter system \cite{Weyl1929,Murakami2007,Wan2011,Balents2011,Huang2015,Weng2015,Nielsen1983,Xu2015a,Xu2015b,Zyuzin2012,Son2013,Liu2013,Liu2014}. The conduction and valence bands of a Weyl semimetal are connected via a number of Weyl points, each of which carries a topological monopole charge given by the surrounding first Chern number $C_1$ in the momentum space \cite{Nielsen1981}. Consequentially, the surface states of a Weyl semimetal take the form of topological Fermi arcs, which connect the projections of oppositely charged Weyl points \cite{Wan2011}. The novel features of Weyl semimetal have motivated the study of other gapless topological phases such as the nodal superconductor \cite{Schnyder2012,Hosur2014}, and their classifications in general \cite{Horava2005,Zhao2013,Schnyder2015,Shunji2013,Yang2014,Yip2014,Chiu2014}.

As an important aspect of study, the generalization of topological states to higher dimensions often reveals profound understandings of topological physics \cite{Zhang2001,Qi2008,Kitaev2009,Ryu2010}. In this work, we present a novel generalization of the Weyl semimetal to five spatial dimensions (5D), which goes beyond the current ten-fold-way classification scheme for gapless topological states \cite{Horava2005,Zhao2013,Schnyder2015}. The Weyl points are generalized to two-dimensional (2D) closed Weyl surfaces, while the topological charge is generalized to the second Chern number $C_2$ of the Berry connection surrounding each Weyl surface in the momentum space. We prove that $C_2$ of a Weyl surface is exactly its topological linking number with other Weyl surfaces in the 5D momentum space. Similar to the 3D Weyl semimetal, this 5D metal hosts topologically protected arcs of Weyl fermions (Weyl fermion arcs) on its four dimensional (4D) boundaries, and the arcs connect between the projections of Weyl surfaces on the boundaries. In cold atom systems and photonic crystals, synthetic spatial dimensions can be created to realize topological phases in higher dimensions \cite{Celi2014,Mancini2015,Stuhl2015,Price2015,Ozawa2016}, which makes the realization of such a 5D topological metal experimentally possible. At last, we briefly discuss the further generalization of Weyl semimetal to generic $2n+1$ spatial dimensions ($n\in\mathbb{Z}^+$), where the corresponding topological charge becomes the $n$-th Chern number $C_n$. The $n$-th Chern number $C_n$ reveals a nontrivial topological relationship among $n$ closed manifolds in $2n+1$ dimensions. These novel metallic states imply the existence of a more complete classification scheme for gapless topological states than the ten-fold way.

\emph{Theoretical formulation.} In this letter, we shall consider crystals that have no symmetries other than the translational symmetry, so that the electron bands are generically nondegenerate in the momentum space except for a few band crossing submanifolds. In the local vicinity of such a band crossing, the Hamiltonian involves only the two crossing bands and therefore must take the following form:
\begin{equation}\label{BC}
H_{\text{cr}}(\mathbf{k})=\xi_0(\mathbf{k})+\xi_1(\mathbf{k})\sigma_{1}+\xi_2(\mathbf{k})\sigma_{2}+\xi_3(\mathbf{k})\sigma_{3}\ ,
\end{equation}
where $\mathbf{k}$ is the momentum, and $\sigma_{i}$ are the Pauli matrices for the orbitals of the two bands. The band crossing submanifold is then given by $\xi_1(\mathbf{k})=\xi_2(\mathbf{k})=\xi_3(\mathbf{k})=0$. In an $n$ dimensional crystal, $\mathbf{k}$ is an $n$-component vector, so the band crossing submanifold is $n-3$ dimensional. Band crossings of this kind are classified as class A in the Altland-Zirnbauer ten-fold-way \cite{Horava2005,Zhao2013,Schnyder2015,Zirnbauer1996,Altland1997}. For $n=3$, this gives exactly the Weyl points in a Weyl semimetal.

The above classification, however, does not capture the global topologies of the band crossing submanifolds for $n>3$ dimensions. On the other hand, in 3D Weyl semimetals, the global topology of a Weyl point is fully characterized by the first Chern number $C_1$ of the Berry connection of the conduction (valence) band  on a 2D surface enclosing the Weyl point. Therefore, it is natural to expect the higher Chern numbers to describe certain global topologies of the band crossings in higher dimensions. In the below, we shall show the minimal generalization is the second Chern number $C_2$ in a class of 5D metals, and show that it is exactly given by the linking number of Weyl surfaces.

We begin by considering the $1$-form U(1) Berry connection of a particular band $|u_\mathbf{k}\rangle$ defined as follows:
\begin{equation}\label{Berry}
A(\mathbf{k})=A_{\mu}(\mathbf{k})\mbox{d}k^{\mu}=i\langle u_{\mathbf{k}}|\partial_{k_\mu}|u_\mathbf{k}\rangle\mbox{d}k^{\mu}\ ,
\end{equation}
where $\mu$ runs over all the spatial dimensions, and repeated indices are summed over. The $2$-form Berry curvature is given by the exterior derivative $F=\mbox{d}A=F_{\mu\nu}\mbox{d}k^\mu\wedge \mbox{d}k^\nu$, where  $F_{\mu\nu}=\partial_{[\mu}A_{\nu]}=(\partial_{\mu}A_{\nu}-\partial_{\nu}A_{\mu})/2$ with $\partial_\mu$ short for $\partial_{k_\mu}$ and $[\ ]$ representing antisymmetrization. Generically, $F$ can be divided into two parts $F=F^{(\ell)}+F^{(t)}$, where $F^{(\ell)}$ and $F^{(t)}$ are the longitudinal and transverse parts satisfying $\mbox{d}^*F^{(\ell)}=0$ and $\mbox{d}F^{(t)}=0$, respectively. Here $\mbox{d}^*=*\mbox{d}*$ is the adjoint of the exterior derivative $\mbox{d}$, and $*$ is the Hodge dual operator. If we regard $A_{\mu}$ as an electromagnetic (EM) vector potential, $F^{(\ell)}$ and $F^{(t)}$ will be produced by monopoles and electric currents, respectively. In particular, the transverse part $F^{(t)}$ has no contribution to the Chern numbers defined on closed manifolds \cite{current}, so we shall simply ignore it and assume $F=F^{(\ell)}$ hereafter.
In 3D, monopoles are point-like particles and are coupled to the $1$-form magnetic field $f$ which is the Hodge dual of the Berry curvature:
\begin{equation}
f=*F=f_\mu \mbox{d}k^\mu, \qquad f_{\mu}=\epsilon_{\mu\nu\lambda}F^{\nu\lambda}\ ,
\end{equation}
where $\epsilon_{\mu\nu\lambda}$ is the Levi-Civita symbol. The condition $\mbox{d}^*F=0$ translates into $\mbox{d}f=0$, i.e., $f$ is curl free, which enables us to define a $0$-form magnetic potential $\phi$ such that $f=\mbox{d}\phi$. For a monopole charge at $\mathbf{k}_0$, $\phi$ satisfies the Poisson equation:
\begin{equation}
\Delta\phi=2\pi\rho_M=\pm2\pi\delta^3(\mathbf{k}-\mathbf{k}_0)\ ,
\end{equation}
where $\Delta=\mbox{d}^*\mbox{d}+\mbox{d}\mbox{d}^*$ is the Laplace operator, $\rho_M$ is the monopole density, and the $2\pi$ factor originates from the Dirac quantization. On a sphere $S^2$ enclosing the monopole, the first Chern number is
\begin{equation}
\begin{split}
&\ C_1=\frac{1}{2\pi}\oint_{S^2} F=\frac{1}{2\pi}\oint_{S^2}*f=\frac{1}{2\pi}\oint_{S^2}*\mbox{d}\phi\\
&=\frac{1}{2\pi}\int_{D^3}\mbox{d}(*\mbox{d}\phi)=\frac{1}{2\pi}\int_{D^3}*\Delta\phi=\pm1\ ,
\end{split}
\end{equation}
where Stokes' theorem is used and $D^3$ is the bulk region enclosed by $S^2$. This is nothing but the Gauss's law for the magnetic field, and the monopoles are exactly the Weyl points that are associated with band $|u_\mathbf{k}\rangle$.

The above picture of EM duality can be immediately generalized to 5D crystals, for which the Hodge dual of the Berry curvature $F$ is a $3$-form:
\[f=*F=f_{\mu\nu\lambda}\mbox{d}k^\mu\wedge\mbox{d}k^\nu\wedge\mbox{d}k^\lambda\ ,\ \ f_{\mu\nu\lambda}=\frac{1}{3!}\epsilon_{\mu\nu\lambda\rho\sigma}F^{\rho\sigma}\ .\]
In this case, monopoles coupled to the dual field $f$ are 2D closed manifolds, or 2-branes in the language of string theory. Physically, these monopoles are just the band crossing submanifolds in the 5D crystal that are associated with band $|u_\mathbf{k}\rangle$. According to Eq. (\ref{BC}), a band crossing submanifold still has a Weyl dispersion in its transverse dimensions, so we shall name it as \emph{Weyl surface} in the following. Again, the condition $\mbox{d}^*F=0$ enables us to rewrite the dual field as $f=\mbox{d}B$, or $f_{\mu\nu\lambda}=\partial_{[\mu}B_{\nu\lambda]}$, where $B$ is a 2-form gauge field. Under the Coulomb gauge $\mbox{d}^*B=\partial^\mu B_{\mu\nu}\mbox{d}k^\nu=0$, the generalized Poisson equation can be written as
\begin{equation}\label{Poisson}
\Delta B^{\mu\nu}=\partial_\lambda\partial^\lambda B^{\mu\nu}=2\pi j^{\mu\nu}\ ,
\end{equation}
where $j^{\mu\nu}$ is the $2$-form monopole density of the Weyl surfaces. For a Weyl surface $\mathcal{M}$ parameterized by 2D coordinates $(\ell_1,\ell_2)$ as $\mathbf{k}_{\mathcal{M}}(\ell_1,\ell_2)$, one can show the monopole density is
\begin{equation}
\begin{split}
&j^{\mu\nu}(\mathbf{k})=\frac{1}{2}\oint_{\mathcal{M}}\mbox{d}\ell_1\mbox{d}\ell_2\ \delta^5\left(\mathbf{k}-\mathbf{k}_{\mathcal{M}}(\ell_1,\ell_2)\right)\\ &\qquad\quad\times\left(\frac{\partial k_{\mathcal{M}}^\mu}{\partial \ell_1}\frac{\partial k_{\mathcal{M}}^\nu}{\partial \ell_2}- \frac{\partial k_{\mathcal{M}}^\nu}{\partial \ell_1}\frac{\partial k_{\mathcal{M}}^\mu}{\partial \ell_2}\right)\\
&=\frac{1}{2}\oint_{\mathcal{M}}\delta^5\left(\mathbf{k}-\mathbf{k}_{\mathcal{M}}\right) \mbox{d}k_{\mathcal{M}}^\mu\wedge\mbox{d}k_{\mathcal{M}}^\nu\ .
\end{split}
\end{equation}
A natural solution to the Poisson equation (\ref{Poisson}) is
\begin{equation}\label{solution}
B^{\mu\nu}(\mathbf{k})=\int \frac{\mbox{d}^5\mathbf{k}'}{4\pi}\frac{j^{\mu\nu}(\mathbf{k}')}{|\mathbf{k}-\mathbf{k}'|^3}=\frac{1}{8\pi}\oint_\mathcal{M} \frac{\mbox{d}k_{\mathcal{M}}^\mu\wedge\mbox{d}k_{\mathcal{M}}^\nu}{|\mathbf{k}-\mathbf{k}_{\mathcal{M}}|^3}\ ,
\end{equation}
and one can easily verify it satisfies the Coulomb gauge.

Now we can investigate the global topology of the Weyl surfaces described by the second Chern number. Unlike a Weyl point in 3D, a Weyl surface alone in 5D can contract itself continuously and vanish identically, and is thus globally trivial. Instead, when there are two 2D Weyl surfaces, they can be nontrivially linked in 5D, where the linking number $L^{(5D)}$ is a global topological invariant \cite{linking}. The simplest example is the Hopf link of two 2D spheres defined by $S^2_a: \{k_1^2+k_2^2+k_3^2=\kappa^2,\ k_4=k_5=0\}$ and $S^2_b:\{ (k_3-\kappa)^2+k_4^2+k_5^2=\kappa^2,\ k_2=k_3=0\}$, respectively, where the linking number is $L^{(5D)}=1$. It is therefore natural to expect the second Chern number $C_2$ to give the linking number of the Weyl surfaces.

Consider two Weyl surfaces $\mathcal{M}_1$ and $\mathcal{M}_2$ associated with band $|u_\mathbf{k}\rangle$ in the 5D momentum space. The gauge field $B$ can then be written as $B=B^{(1)}+B^{(2)}$, where $B^{(1)}$ and $B^{(2)}$ are the solutions to the two Weyl surfaces as given in Eq. (\ref{solution}), respectively. We can draw a 4D closed manifold $\partial\mathcal{V}$ being the boundary of a 5D region $\mathcal{V}$ that encloses $\mathcal{M}_1$ but not $\mathcal{M}_2$. This can always be done by choosing $\mathcal{V}$ in the vicinity of $\mathcal{M}_1$ and thin enough in the transverse dimensions of the $\mathcal{M}_1$ surface. This is analogous to the 3D case, when given two loops $\mathcal{L}_1$ and $\mathcal{L}_2$ linked together, one can always draw a thin torus $\partial\mathcal{V}$ around $\mathcal{L}_1$ that encloses $\mathcal{L}_1$ but not $\mathcal{L}_2$.  We then define the second Chern number $C_2$ of the Weyl surface $\mathcal{M}_1$ as
\begin{equation}
C_2(\mathcal{M}_1)=\frac{1}{8\pi^2}\oint_{\partial\mathcal{V}}F\wedge F\ .
\end{equation}
By noting $F=*f=*\mbox{d}B$ and using the Stokes' theorem, we can rewrite $C_2(\mathcal{M}_1)$ as
\begin{widetext}
\begin{equation}\label{C2}
C_2(\mathcal{M}_1)=\oint_{\partial\mathcal{V}}\frac{*\mbox{d}B\wedge *\mbox{d}B}{8\pi^2}=\int_{\mathcal{V}}\frac{\mbox{d}(*\mbox{d}B\wedge *\mbox{d}B)}{8\pi^2} =\int_{\mathcal{V}}\frac{\mbox{d}^*\mbox{d}B\wedge \mbox{d}B}{4\pi^2}
=\int_{\mathcal{V}}\frac{\Delta B\wedge \mbox{d}B}{4\pi^2} =\int_{\mathcal{V}}\frac{\Delta B^{(1)}\wedge \mbox{d}B^{(2)}}{4\pi^2}=L^{(5D)}_{\mathcal{M}_1}\ ,
\end{equation}
where
\begin{equation}\label{link}
L^{(5D)}_{\mathcal{M}_1}=\frac{1}{(2!)^2}\frac{3}{8\pi^2}\oint_{\mathcal{M}_1}\mbox{d}k_{\mathcal{M}_1}^\mu\wedge\mbox{d}k_{\mathcal{M}_1}^\nu \oint_{\mathcal{M}_2}\mbox{d}k_{\mathcal{M}_2}^\lambda\wedge\mbox{d}k_{\mathcal{M}_2}^\rho \frac{\epsilon_{\mu\nu\lambda\rho\sigma}(k_{\mathcal{M}_1}^\sigma-k_{\mathcal{M}_2}^\sigma)} {|\mathbf{k}_{\mathcal{M}_1}-\mathbf{k}_{\mathcal{M}_2}|^5}
\end{equation}
\end{widetext}
is exactly the linking number between $\mathcal{M}_1$ and $\mathcal{M}_2$ as we expected. In deriving Eq. (\ref{C2}), we have used the facts $\Delta B^{(1)}\wedge \mbox{d}B^{(1)}=0$ and $\Delta B^{(2)}=0$ inside the region $\mathcal{V}$. The expression in Eq. (\ref{link}) is simply a 5D generalization of the Gauss linking number of loops in 3D given by Polyakov \cite{Wilczek1983,Polyakov1988,Witten1989}. The geometrical meaning of Eq. (\ref{link}) is the total solid angle wrapped by $(\mathbf{k}_{\mathcal{M}_1}-\mathbf{k}_{\mathcal{M}_2})$ when $\mathbf{k}_{\mathcal{M}_1}$ and $\mathbf{k}_{\mathcal{M}_2}$ runs over $\mathcal{M}_1$ and $\mathcal{M}_2$, respectively, divided by the total solid angle of a 4D sphere, $8\pi^2/3$. Note the sign of $L^{(5D)}_{\mathcal{M}_1}$ depends on the orientation of the momentum space.

When there are $N>2$ Weyl surfaces $\mathcal{M}_j$ ($1\le j\le N$) associated with the same band $|u_\mathbf{k}\rangle$, it is straightforward to prove that $C_2(\mathcal{M}_j)=L^{(5D)}_{\mathcal{M}_j}$ is the total linking number of $\mathcal{M}_j$ with the other Weyl surfaces. Since the linking number in Eq. (\ref{link}) reverses sign under the interchange of $\mathcal{M}_1$ and $\mathcal{M}_2$, one would always have $\sum_{j=1}^NC_2(\mathcal{M}_j)=0$. In analogy to the 3D Weyl semimetal where the Weyl-point first Chern number $C_1$ leads to surface fermi arcs, the second Chern number $C_2(\mathcal{M}_j)$ also produces Weyl fermion arcs on the 4D boundary of the 5D metal, as we shall show in the explicit model constructed below.

\emph{An explicit model}. To verify the theory and the prediction we derived above, it is instructive to construct a lattice model of the 5D topological metal. A simple example is a four-band model Hamiltonian as given below:
\begin{equation}\label{Hk}
H(\mathbf{k})=\sum_{i=1}^5\zeta_i(\mathbf{k})\Gamma^i+ia\frac{[\Gamma_4,\Gamma_5]}{2}\ ,
\end{equation}
where $\Gamma^i$ ($1\le i\le 5$) are the five $4\times 4$ Gamma matrices satisfying the anticommutation relation $\{\Gamma^i,\Gamma^j\}=2\delta^{ij}$, the functions $\zeta_i(\mathbf{k})$ are defined as $\zeta_1(\mathbf{k})=\sin k_1$, $\zeta_2(\mathbf{k})=\sin k_2$, $\zeta_3(\mathbf{k})=\sin k_3+w(3-\cos k_1-\cos k_2-\cos k_4)$, $\zeta_4(\mathbf{k})=\sin k_4+t(1-\cos k_5)$, $\zeta_5(\mathbf{k})=\sin k_5$, and $a$ is a number satisfying $0<a<1$. Hereafter we shall take the Gamma matrix representation $\Gamma^{1,2,3,4,5}=\{\sigma_3\tau_1,\ \sigma_3\tau_2,\ \sigma_3\tau_3,\ \sigma_1,\ \sigma_2\}$, where $\sigma_{1,2,3}$ and $\tau_{1,2,3}$ are both Pauli matrices. The parameters in the model are constrained by $2t>1$ and $2w>a+1$. The dispersions of the four bands of $H(\mathbf{k})$ can be easily derived to be
\begin{equation}\label{Ek}
\epsilon_{i}(\mathbf{k})=\nu_i\sqrt{(\eta(\mathbf{k})+\lambda_i a)^2+\zeta_4(\mathbf{k})^2+\zeta_5(\mathbf{k})^2}\ ,
\end{equation}
where $\epsilon_{i}(\mathbf{k})$ ($i=1,2,3,4$) stands for the $i$-th lowest band, $\eta(\mathbf{k})=\sqrt{\zeta_1(\mathbf{k})^2+\zeta_2(\mathbf{k})^2+\zeta_3(\mathbf{k})^2}$, while $\nu_i$ and $\lambda_i$ are signs defined as $\nu_i=(-,-,+,+)$ and $\lambda_i=(+,-,-,+)$. We shall focus on the third band $\epsilon_3(\mathbf{k})$ to see whether Eq. (\ref{C2}) holds.

\begin{figure}
\begin{center}
\includegraphics[width=3.3in]{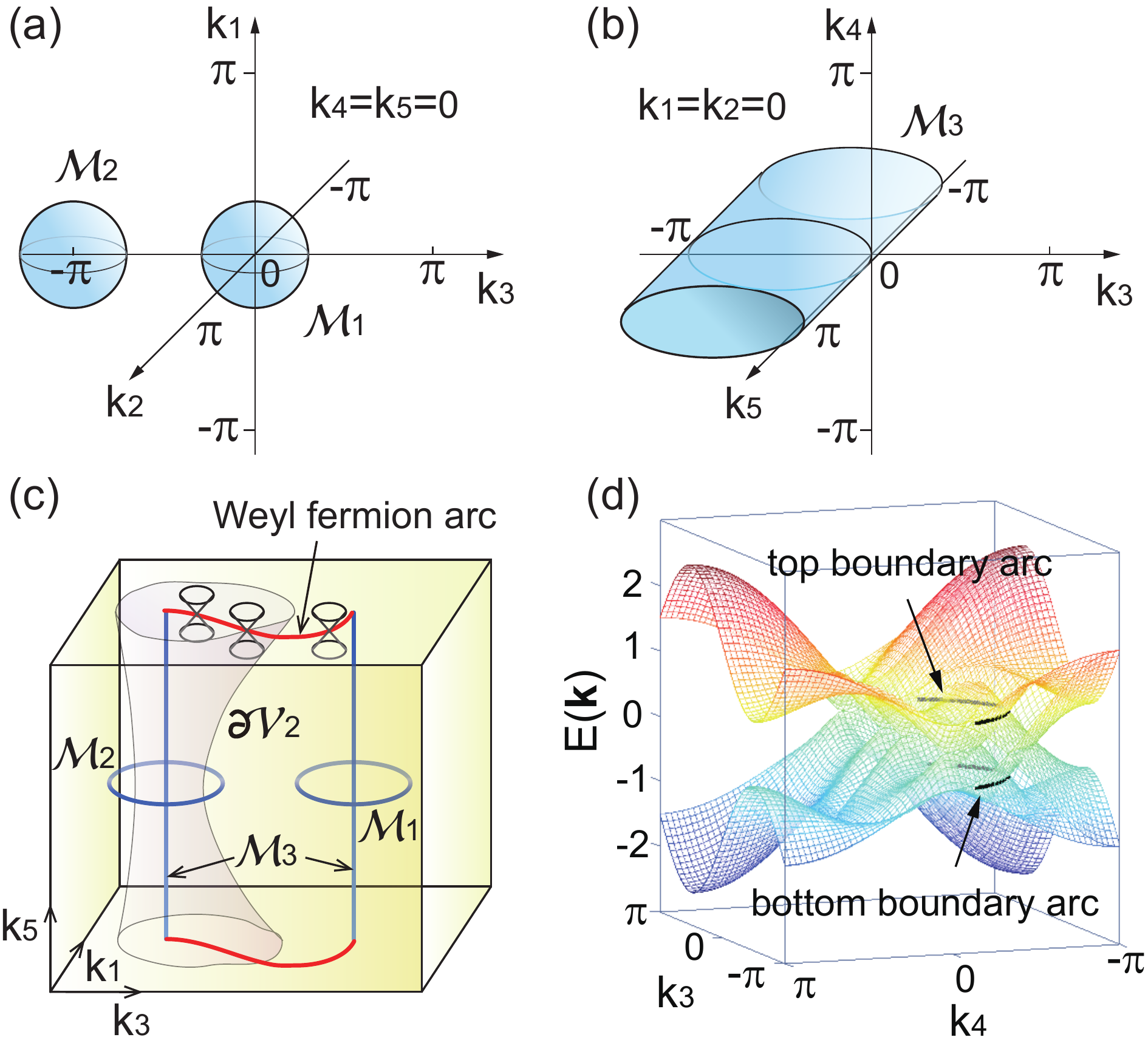}
\end{center}
\caption{(color online). (a) Weyl surfaces $\mathcal{M}_1$ and $\mathcal{M}_2$ plotted in the subspace $k_4=k_5=0$. (b) Weyl surface $\mathcal{M}_3$ plotted in the subspace $k_1=k_2=0$. (c) The Weyl surfaces can be seen linked in the subspace $k_2=k_4=0$. A Weyl fermion arc occurs on the boundary and crosses all the possible $\partial \mathcal{V}_2$. (d) Energy spectrum $E(\mathbf{k})$ in the subspace $k_1=k_2=0$ calculated with open boundary condition in the $k_5$ direction, where there are two arcs from the top and bottom boundaries, respectively.}
\label{5Dlinking}
\end{figure}

From the energy dispersion in Eq. (\ref{Ek}), it is straightforward to show that band $\epsilon_3(\mathbf{k})$ is associated with three Weyl surfaces in the momentum space: the first two Weyl surfaces $\mathcal{M}_{1,2}$ are between bands $\epsilon_2(\mathbf{k})$ and $\epsilon_{3}(\mathbf{k})$, which are given by $\eta(\mathbf{k})-a=\zeta_4(\mathbf{k})=\zeta_5(\mathbf{k})=0$. Topologically, they are both 2D spheres in the subspace $k_4=k_5=0$ of the Brillouin zone as shown in Fig. \ref{5Dlinking}(a). The third Weyl surface $\mathcal{M}_3$ is between bands $\epsilon_3(\mathbf{k})$ and $\epsilon_{4}(\mathbf{k})$, and is described by $\zeta_1(\mathbf{k})=\zeta_2(\mathbf{k})=\zeta_3(\mathbf{k})=0$. As is plotted in Fig. \ref{5Dlinking}(b), $\mathcal{M}_3$ is a 2D torus in the subspace $k_1=k_2=0$. Both of the two spheres $\mathcal{M}_1$ and $\mathcal{M}_2$ are topologically linked with the torus $\mathcal{M}_3$. One way to see this is to plot $\mathcal{M}_{1,2,3}$ in the $k_2=k_4=0$ slice of the momentum space, where they appear as linked loops as shown in Fig. \ref{5Dlinking}(c).

Now we can calculate the second Chern numbers of the Weyl surfaces using the Berry connection of band $\epsilon_3(\mathbf{k})$. The wave function of band $\epsilon_3(\mathbf{k})$ can be written as
\begin{equation}
\begin{split}
&|u_{3,\mathbf{k}}\rangle = \Big(\cos\frac{\alpha}{2}\cos\frac{\theta}{2},\ \cos\frac{\alpha}{2}\sin\frac{\theta}{2}e^{i\varphi},\\
&\quad\quad \sin\frac{\alpha}{2}\cos\frac{\theta}{2}e^{i\psi},\ \sin\frac{\alpha}{2}\sin\frac{\theta}{2}e^{i\psi+i\varphi}\Big)^T\ ,
\end{split}
\end{equation}
where the angles $\theta,\ \varphi,\ \alpha$ and $\beta$ are defined by $\cos\theta=\zeta_3(\mathbf{k})/\eta(\mathbf{k})\ $, $e^{i\varphi}\sin\theta =[\zeta_1(\mathbf{k})+i\zeta_2(\mathbf{k})]/\eta(\mathbf{k})\ $, $\cos\alpha=[\eta(\mathbf{k})-a]/\epsilon_3(\mathbf{k})$ and $e^{i\psi}\sin\alpha =[\zeta_4(\mathbf{k})+i\zeta_5(\mathbf{k})]/\epsilon_3(\mathbf{k})\ $. Therefore, the $1$-form Berry connection of band $\epsilon_3(\mathbf{k})$ takes the form
\begin{equation}
A=i\langle u_{3,\mathbf{k}}|\mbox{d}|u_{3,\mathbf{k}}\rangle=\frac{\cos\theta-1}{2}\mbox{d}\varphi+\frac{\cos\alpha-1}{2}\mbox{d}\psi\ ,
\end{equation}
from which we find $F\wedge F=(1/2)\sin\theta\sin\alpha\ \mbox{d}\theta\wedge\mbox{d}\varphi\wedge\mbox{d}\alpha\wedge\mbox{d}\psi$. The next step is to construct a 4D manifold that encloses the Weyl surface we are interested in. Since $\epsilon_3(\mathbf{k})\ge0$ and reaches zero only on $\mathcal{M}_1$ and $\mathcal{M}_2$, the Fermi surface at $\epsilon_3(\mathbf{k})=\epsilon_F$ with  $\epsilon_F$ positive and sufficiently small ($0<\epsilon_F<1-a$) naturally consists of two 4D manifolds $\partial\mathcal{V}_1$ and $\partial\mathcal{V}_2$ that enclose $\mathcal{M}_1$ and $\mathcal{M}_2$, respectively. Both $\partial\mathcal{V}_1$ and $\partial\mathcal{V}_2$ have a topology $S^2\times S^2$ where $S^2$ stands for a 2D sphere, and the angles $\theta,\varphi$ and $\alpha,\psi$ wind exactly once around the former and latter $S^2$ in the direct product, respectively. While the winding orientations of $\alpha,\psi$ on $\partial\mathcal{V}_1$ and $\partial\mathcal{V}_2$ are the same, the winding orientations of $\theta, \varphi$ on them are opposite. As a result, the second Chern numbers of $\mathcal{M}_{1}$ and $\mathcal{M}_2$ are
\begin{equation}
\begin{split}
&C_2(\mathcal{M}_{1,2})=\frac{1}{8\pi^2}\oint_{\partial\mathcal{V}_{1,2}}F\wedge F=\pm\int_{0}^\pi\mbox{d}\theta\int_{0}^{2\pi}\mbox{d}\varphi\\
&\times\int_{0}^\pi\mbox{d}\alpha\int_{0}^{2\pi}\mbox{d}\psi \frac{\sin\theta\sin\alpha}{16\pi^2}=\pm1\ ,
\end{split}
\end{equation}
respectively. Similarly, one can show the second Chern number of the third Weyl surface $\mathcal{M}_3$ is zero. This is exactly what we expect from Eq. (\ref{C2}) and the topological linking numbers of the Weyl surfaces.

The second Chern number of the Weyl surfaces will lead to a Weyl fermion arc on the boundary of the 5D metal. Heuristically, we can think of the 4D manifold $\partial\mathcal{V}_{2}$ surrounding $\mathcal{M}_2$ as a 4D gapped system with second Chern number $C_2=-1$. Such a 4D system is known to exhibit the 4D quantum Hall effect and host surface states in the form of a (3+1)D Weyl fermion \cite{Zhang2001,Qi2008}. Furthermore, $\partial\mathcal{V}_{2}$ can be deformed freely as long as it does not touch any Weyl surfaces. Therefore, for instance, if we take an open boundary condition in the $k_5$ direction,  we will obtain a 1D arc of (3+1)D Weyl fermion on the boundary that crosses any possible $\partial\mathcal{V}_{2}$, as illustrated in the $k_2=k_4=0$ momentum space slice shown in Fig. \ref{5Dlinking}(c). The projection of $\mathcal{M}_3$ on the boundary is then connected by the arc.

In general, such a Weyl fermion arc can be verified easily by numerical calculations, but is hard to visualize as a high dimensional object. Here in our model, however, a symmetry $(\Gamma^1\Gamma^2)^\dag H(k_1,k_2,k_3,k_4,k_5)\Gamma^1\Gamma^2=H(-k_1,-k_2,k_3,k_4,k_5)$ enables us to see the arc in a lower dimensional subspace. This symmetry ensures the energy spectrum to be symmetric under the partial inversion $(k_1,k_2)\rightarrow(-k_1,-k_2)$, so the only Weyl fermion arc must be lying in the $k_1=k_2=0$ subspace. Fig. \ref{5Dlinking}(d) shows the energy bands calculated with an open boundary condition in the $k_5$ direction and $k_1=k_2=0$, and one can find two arcs coming from the top and bottom 4D boundaries, respectively. Due to another symmetry $\Gamma^{5\dag} H(k_1,k_2,k_3,k_4,k_5)\Gamma^5=-H(k_1,k_2,k_3,k_4,-k_5)$, the two arcs on the top and bottom boundaries have opposite energies, which are related to each other by $k_5\rightarrow -k_5$.

\emph{Higher dimensional generalizations.} This scheme of generalizing the Weyl semimetal can be continued to arbitrary $2n+1$ dimensional solids ($n\in\mathbb{Z}^+$), where the band crossing submanifolds $\mathcal{M}_j$ are $2n-2$ dimensional. Similarly, we can define the $n$-th Chern number $C_n$ of a band crossing $\mathcal{M}_1$ on a $2n$ dimensional manifold $\partial\mathcal{V}$ that encloses $\mathcal{M}_1$. When there are $n$ band crossings $\mathcal{M}_j$ ($1\le j\le n$), $C_n$ can be rewritten as
\begin{equation}
\begin{split}
&C_n(\mathcal{M}_1)=\oint_{\partial\mathcal{V}}\frac{F^n}{n!(2\pi)^n}=\oint_{\mathcal{M}_1}\prod_{r=2}^{n}\wedge \Big[\frac{\mbox{d}k_{\mathcal{M}_1}^{\mu^r_{1}}\wedge\mbox{d}k_{\mathcal{M}_1}^{\mu^r_{2}}}{2!(2n-2)!\Omega_{2n}}\\
&\times\oint_{\mathcal{M}_r}\frac{\epsilon_{\mu^r_1\cdots\mu^r_n}\left(k_{\mathcal{M}_1}^{\mu^r_{3}}-k_{\mathcal{M}_r}^{\mu^r_{3}}\right) \mbox{d}k_{\mathcal{M}_r}^{\mu^r_{4}}\wedge\cdots\wedge\mbox{d} k_{\mathcal{M}_r}^{\mu^r_{2n+1}}} {|\mathbf{k}_{\mathcal{M}_1}-\mathbf{k_{\mathcal{M}_r}}|^{2n+1}}\Big]\ ,
\end{split}
\end{equation}
where $\Omega_{2n}$ is the area of the $2n$ dimensional unit sphere. It characterizes a certain topology between the $n$ band crossings, which is, however, not yet fully understood at this moment. Such a topological number may play a role in the understandings of high dimensional knot theory, which may have applications in string theory and other high dimensional theories.

\emph{Discussions.} The above 5D generalization of Weyl semimetals based on the second Chern number $C_2$ can be viewed as a finer classification in 5D for gapless phases of class A that have no symmetries. This indicates that the previous classification scheme of noninteracting gapless phases based on the Altland-Zirnbauer ten-fold-way \cite{Horava2005,Zhao2013,Schnyder2015} is not yet complete in high dimensions. Our work shed light on a more complete classification which could be carried out in the future. Besides, the recent development of synthetic dimensions makes it possible to implement additional spatial dimensions with certain internal degrees of freedoms in cold atom systems or photonic crystals \cite{Celi2014,Mancini2015,Stuhl2015,Price2015,Ozawa2016}. This could possibly lead to the realization of such a 5D topological metal in experiments. Finally, the topological meaning of the $n$-th Chern number $C_n$ in $2n+1$ spatial dimensions is yet to be made clear. With the band crossing manifolds resembling the D-branes in the superstring theory, the high dimensional knot structure indicated by $C_n$ together with the high dimensional fermi arcs on the boundaries may have potential reinterpretations and applications in string theories or high energy theories with extra dimensions \cite{Horava2005,Ishikawa1995,Strominger1996}.

\emph{Acknowledgement.} We thank Quan Zhou and Chao-Ming Jian for helpful discussions. This work is supported by the NSF under grant number DMR-1305677.

\bibliography{5DWeyl_ref}

\begin{thebibliography}{45}
\bibitem[{cur()}]{current}
\bibinfo{note}{Calculation of the ($n$-th) Chern number only involves
  integrations of $F^{(t)}$ on closed manifolds, which are always zero since
  $\mbox{d}F^{(t)}=0$ in the entire momentum space.}
\bibitem[{lin()}]{linking}
\bibinfo{note}{Two closed manifolds of dimensions $d_1$ and $d_2$ respectively
  can be nontrivially linked in a $d_1+d_2+1$ dimensional space.}
\end{thebibliography}

\end{document}